\documentstyle[multicol,prl,aps]{revtex}
\input psfig
\pssilent
\begin{document}
\title{ Making classical and quantum canonical general relativity
computable through a power series expansion in the inverse
cosmological constant } 

\author{ Rodolfo Gambini$^{1}$, Jorge Pullin$^2$} 

\address{1. Instituto de F\'{\i}sica, Facultad de
Ciencias, Igu\'a 4225, esq. Mataojo, Montevideo, Uruguay.}  

\address{
2. Center for Gravitational Physics and Geometry, Department of
Physics,\\ The Pennsylvania State University, 104 Davey Lab,
University Park, PA 16802.}  \date{August 14th 2000} \maketitle

\begin{abstract}
We consider general relativity with a cosmological constant as a
perturbative expansion around a completely solvable diffeomorphism
invariant field theory. This theory is the $\Lambda\rightarrow\infty$
limit of general relativity. This allows an explicit perturbative
computational setup in which the quantum states of the theory and the
classical observables can be explicitly computed. The zeroth order
corresponds to highly degenerate space-times with vanishing
volume. Perturbations give rise to space-times with non-vanishing
volumes in a natural way.  The spectrum of area- and volume-related
observables constructed by coupling the theory to matter can be
directly assessed.  An unexpected relationship arises at a quantum
level between the discrete spectrum of the volume operator and the
allowed values of the cosmological constant.
\end{abstract}
\begin{multicols}{2}

The recent years have been very fertile for the development of
canonical quantizations of general relativity. There are now available
more than one \cite{QSD,Vassil} highly non-trivial mathematically
consistent, finite, anomaly-free theories of canonical quantum
gravity. Instrumental in these developments have been the underlying
advances in mathematical techniques for dealing with infinite
dimensional nonlinear spaces, like the theory of cylindrical functions
and associated measures \cite{AsLe}, and the introduction of spin
networks to eliminate the over-completeness of the Wilson loop basis
\cite{RoSmvo} (for a recent summary, see the review article by Rovelli
\cite{Roliving}).  In spite of the advances, most of the results
obtained from these theories up to now concern statements made {\em at
a kinematical level} (without imposing the Hamiltonian
constraint). One of the remaining main challenges is to obtain results
that hold at a dynamical level, imposing the full set of constraints
of the theory. This requires making the theories computable, in the
sense of obtaining classical observables to promote to quantum
observables and solving for the physical set of states that are
annihilated by all the constraints. Only in this setting will one be
able to make contact with trustworthy physical predictions.

The fact that implementing the full dynamics of canonical quantum
gravity is hard to do has plagued the subject since its inception in
the 60's by DeWitt\cite{DeWitt}.  For instance, one does not know even
at a classical level how to compute observable quantities for
canonical general relativity (quantities that are invariants under the
symmetries of the theory or ---in the canonical language--- that
commute with the constraints), and it is known that such quantities
are potentially quite involved \cite{To}. As a consequence, it is very
hard to make contact with the semiclassical limit, since one does not
have at hand observable quantities to compute. Moreover, the space of
quantum states that are annihilated by the quantum constraints cannot
be straightforwardly computed \cite{Smstate}. None of these issues was
directly tackled by the significant amount of progress that was made
in terms of the introduction of connection type variables and loop
representations for quantum gravity that we mentioned at the
beginning. There is general consensus that an approximation scheme is
desirable in such a way as to make the above problems solvable.  The
challenge consists in finding an approximation scheme that will not
conflict with non-perturbative nature of the canonical quantization,
in particular the diffeomorphism invariance of the theory.

In this paper we would like to propose a computational scheme that
allows to explicitly address the issue of finding observables for the
theory (at a classical level) and to find quantum states annihilated
by all the constraints. The central idea will be to consider general
relativity as a power series expansion in terms of the inverse
cosmological constant, and to view the full theory as a
``perturbation'' of a zeroth order solvable theory. The zeroth order
theory corresponds to general relativity in the limit in which the
cosmological constant $\Lambda\rightarrow\infty$. The use of an
approximation based on a large cosmological constant has been
pioneered in a different context (that of studying the spectrum of the
physical Hamiltonian obtained by ``de-parameterizing'' the theory) by
Smolin \cite{Smbrill} and also by Rovelli \cite{RoT}, with some level
of success. We will however be able to make further progress, in
significant part due to the new maturity of the field we mentioned
above but also because we push the perturbative approach at all
levels: constraints, observables and equations of motion, both
classically and quantum mechanically. It might be argued that a scheme
where the cosmological constant is large (in Planck units) lacks
physical relevance, given is small present value. However, having a
regime of the theory in which it is completely solvable can be a
valuable tool, even if such regime is unphysical. Moreover, as we will
discuss below, there is some evidence that our approach might be
useful beyond its apparent domain of validity.  It might be the case
that the situation is analogous to the large $N$ approximation in QCD
or the strong coupling expansions of lattice field theory, that 
a priori look less promising that what they end up being.

The starting point is to consider canonical quantum gravity with a
cosmological constant. The Hamiltonian constraint of the theory is,
\begin{equation}
H(N)=H_0(N)+\Lambda \int d^3 x N(x) \sqrt{{\rm det} q(x)}
\end{equation}
where $N(x)$ is a smearing function, $H_0(N)$ is the Hamiltonian
constraint without a cosmological constant and ${\rm det} q$ is the
determinant of the spatial metric. If one now considers the limit
$\Lambda\rightarrow\infty$, and re-scales the constraint by 
$1/\Lambda$
\cite{HuKu}
one is left with a theory, which we will call ``zeroth order'' theory
for which the Hamiltonian constraint is,
$
%\begin{equation}
H^{(0)}(N)=\int d^3x N(x) \sqrt{{\rm det} q(x)},
%\end{equation}
$
that is, the Hamiltonian constraint is just the square root of the
determinant of the spatial metric. In addition to this the theory has
the ordinary diffeomorphism constraint and if one uses Ashtekar
variables 
there will also be a Gauss law constraint (both are independent of
$\Lambda$). Imposing classically the Hamiltonian constraint of the
zeroth order theory, one immediately sees that it corresponds to
metrics of identically vanishing determinant, that is, degenerate
metrics.

The dynamics of the zeroth order theory is easily solved.  One can
readily construct quantities that have vanishing Poisson bracket with
the Hamiltonian constraint of the zeroth order theory. For instance,
one can consider any function depending only on the three metric (or
if one is using Ashtekar variables $(\tilde{E}^a_i, A_b^j)$, the
densitized triads $\tilde{E}^a_i$) and not on its canonically
conjugate momenta. To be a genuine observable the quantity should
also have vanishing Poisson bracket with the diffeomorphism
constraint. A way of constructing quantities with this property is to
couple the theory with matter fields and physically define geometric
objects like surfaces or volumetric regions through properties of the
matter fields \cite{obs,Smobs} and then compute the areas or volumes
of these regions. The usual problem with this construction is that the
addition of matter adds terms to the Hamiltonian constraint as well,
and the areas and volumes computed as we just described do not commute
with the Hamiltonian constraint.  However, in the
$\Lambda\rightarrow\infty$ limit the matter terms drop out of the
Hamiltonian and one ends up (in the zeroth order theory) with genuine
observables (this is similar to what happens in models without a
Hamiltonian constraint as was discussed by Rovelli in the
Husain--Kucha\v{r} \cite{HuKu} model \cite{RoT}).  A concrete example
of an observable for the zeroth order theory is obtained by coupling
the theory to an antisymmetric tensor field with the standard
Kalb--Ramond free Hamiltonian as proposed by Smolin \cite{Smobs}. The
antisymmetric tensor field is naturally associated with surfaces of
which one can compute the area. The resulting quantity, which
essentially is a contraction of the triads with the canonically
conjugate momenta of of the Kalb--Ramond field, is naturally
diffeomorphism invariant. One can build a ``loop-surface''
representation (see \cite{Smobs,pio} for details) for the coupled
theory by expanding the gauge invariant wavefunctions in a basis of
states $|\Sigma,s>$ labeled by equivalence classes under
diffeomorphisms of spin networks $s$ and three-surfaces $\Sigma$. A
natural generalization of the Ashtekar-Lewandowski \cite{AsLe} measure
allows to introduce an inner product. On this basis the area defined
by the fields has a discrete spectrum.

The Hamiltonian constraint of the zeroth order theory can be readily
realized in the kinematical space \cite{RoSmvo,AsLevo} of
non-diffeomorphism invariant wavefunctions  as,
\begin{equation}
H^{(0)}(N) |\Sigma,s> = \lim_{\epsilon\rightarrow 0} 
\sum_{v\in s} N(v) V(v,\epsilon) |\Sigma,s>
\end{equation}
where we have represented, following Thiemann \cite{QSD}, the
determinant of the metric as the volume operator of an infinitesimal
region of size $\epsilon$ centered on the vertices $v$ of the spin
network. The operator is independent on $\epsilon$ on this space of
states and therefore the above limit is well defined. 
To consider the action on diffeomorphism invariant states one 
evaluates the inner product $<\{\Sigma'\},\{s\}|H^{(0)}(N)
|\Sigma,s>$ which is well defined. We can readily find 
the space of states that are annihilated by the Hamiltonian
constraint of the zeroth order theory: they correspond to states based
on spin networks such that the localized volume operator vanishes on
each of their vertices. For trivalent vertices this is the case for
all states. For four and higher valent vertices the operator vanishes
automatically if the vertices are planar (this is not true for the
volume operator of \cite{RoSmvo}, but it holds for the definition
of \cite{AsLevo}). If the vertices are not
planar it is also possible to find values of the valences of the
incoming edges such that the operator vanishes. We collectively denote
all these states as $|\Sigma,s^0>$. Since the space of
solutions of the Hamiltonian constraint is a subspace of the states we
were considering, it can be readily endowed with an inner product
which is just the restriction to this space of the inner product
derived from the Ashtekar--Lewandowski measure. Notice that the area
observables we introduced are well defined on this space.

We have therefore seen that the zeroth order theory is well
defined. It is non-perturbatively quantized in a basis of spin
networks and surfaces, one can solve for the physical states,
construct observables, and introduce a physical inner product. This is
therefore a sound starting point for a perturbative expansion. The
resulting perturbative theory is remarkably similar to bound state
perturbation theory in quantum mechanics in the degenerate case.  The
operator $H^{(0)}$ (which can be readily diagonalized in the basis of
states considered) has a degenerate spectrum.

To consider the higher order corrections to the zeroth order theory,
one is interested in solving the eigenvalue problem
\begin{equation}
<\phi(\Lambda^{-1})| H^{(0)} +\Lambda^{-1} H^{(1)} =
\epsilon(\Lambda) <\phi(\Lambda^{-1}| \label{hlambda}
\end{equation}
with eigenvalue $\epsilon(\Lambda)$ equal to zero up to the desired
order in perturbation theory. We will assume that the states and the
eigenvalue can be expanded in power series in the inverse cosmological
constant, and address the resulting equations order by order. Since
the spectrum of the zeroth order Hamiltonian is degenerate, we apply
degenerate perturbation theory to solve this problem. One starts by
considering a state with eigenvalue $\epsilon^{(0)}$ for the zeroth
order Hamiltonian constraint, that is, a spin network state with a
definite volume. Notice that this is {\em not} a state of the zeroth
order theory. Considering first order corrections and solving the
equations resulting from (\ref{hlambda}) one determines the first
order corrections and a polynomial $\epsilon(\Lambda^{-1})$ which, set
to zero, determines the value of $\Lambda$ for which the perturbative
treatment is valid. That is, given an eigenvalue of the zeroth order
Hamiltonian, one constructs a solution of the first order theory that
has vanishing Hamiltonian for a particular value of lambda. That is,
as a consequence of the discrete spectrum of the zeroth order
Hamiltonian inherited from the discreteness of the volume operator,
the cosmological constant is, in this approach, quantized.

This might sound unusual, but imposing the constraint perturbatively
is analogous to considering an atom in a magnetic field and trying to
find the states with a given energy level. Such energy level can be
reached by choosing a nearby energy state of an unperturbed atom and
then tuning the magnetic field to a given value.  Various states could
correspond to the same energy, depending on the value of the magnetic
field. The approximate value of the magnetic field required will also
depend on how high one goes in perturbation theory.  A detailed
discussion of how this is implemented in several systems
can be seen in \cite{GaPu00b}.

A remarkable aspect is that all the calculations involved in the
perturbative approach are completely explicit when one works in the
spin network basis. One starts by considering eigenvalues and
eigenvectors of the zeroth order Hamiltonian. Since this Hamiltonian
is the volume operator, which is well understood, the space of states
is under control (even when it is non-vanishing). The equation that
determines the first order correction involves evaluating the
Hamiltonian constraint on the states with a given volume. Such
calculations are completely explicit given, for instance, the
Hamiltonian introduced by Thiemann \cite{QSD}, or the Hamiltonian in
the Vassiliev arena \cite{Vassil}. The only subtle element in the
calculation is that given the infinite degeneracy of the volume
operator, the first order correction might involve infinite
superpositions and therefore may lead us to corrections to the
kinematical inner product. It is worthwhile noticing that in general
the first order corrections have non-zero volume, even if one started
from the zeroth order state with vanishing volume, as long as one did
not restrict to either planar or trivalent spin networks. So the
resulting theory is a theory of non-vanishing volume space-times.

The above construction has a classical counterpart when one tries to
find observables for the theory. An observable for the first order
theory can be written as a perturbation of the observables of the
zeroth order level theory we discussed above,
\begin{equation}
O_\Lambda(\pi,\tilde{E})=O^{(0)}(\pi,\tilde{E})+\Lambda^{-1}
O^{(1)}(\pi,\tilde{E}).
\end{equation}
One would like to request that these observables have vanishing
Poisson brackets with the Hamiltonian of the theory (more precisely,
weakly vanishing, but the extension to such case is immediate). 
Expanding such
requirement in powers of $\Lambda^{-1}$ one gets to zeroth and first
order in $\Lambda^{-1}$,
\begin{eqnarray}
\left\{O^{(0)},H^{(0)}\right\}&=&0\\
\left\{O^{(1)},H^{(0)}\right\}+\left\{O^{(0)},H^{(1)}\right\}&=&0
\label{seconcon}
\end{eqnarray}

The first equation determines $O^{(0)}$ and the second one leads to a
(functional) linear partial differential equation for $O^{(1)}$. The
construction can be readily continued to higher orders. In all cases
one obtains a linear partial differential equation, albeit with a more
and more complex inhomogeneous term. It should be noted that one can
obtain many observables starting with different solutions to the first
equation. The linear partial differential equations are not hard to
solve, given that the coefficients of the derivatives are functions of
the triads, whereas the derivatives are with respect to the
connections. Given its simplicity, the system is always integrable and
therefore yields all the observables of the theory.  The solutions to
these equations will generically be ill defined or non-convergent in
certain regions of phase space. This is expected in theories with
complicated dynamics and is not a pathology of the method per se but
rather a feature of the observable quantities.  In particular, one
might be concerned that the current approach produces {\em too many}
observables. For instance, there exist pathological systems that have
less observables than degrees of freedom (e.g. the heavy asymmetric
top). A priori our method seems to yield in these cases extra
observables, but on closer examination the quantities constructed are
not acceptable (in the case of the top they take values on a
restricted portion of phase space not preserved by evolution). We
discuss several pathological examples and how the method correctly
handles them in \cite{GaPu00b}.

Let us consider, as an simple example of interest in gravity, the
computation of an observable for  Bianchi cosmologies. Observables
for these models (with cosmological constant) had never been studied
before. We will write the Bianchi models in terms of Ashtekar's
variables following the notation of \cite{AsPu}, in which one codes
all the information in three variables associated with the time
dependent portion of the triads $E^1,E^2,E^3$ and their conjugate
momenta $A_1,A_2,A_3$. The only constraint of the theory given the
homogeneity and diagonality of the metric is the Hamiltonian, which
rescaled by the cosmological constant reads,
\begin{equation}
H= E^1 E^2 E^3 + \Lambda^{-1} \left(
E^1 E^2 (A_1 A_2 -\epsilon A_3) +{\rm cyclic}
\right), 
\end{equation}
where $\epsilon=1$ corresponds to the Bianchi IX model and
$\epsilon=0$ to the Bianchi I model and the extra terms are obtained
by cyclically permuting the indices $1,2,3$. It is straightforward
to repeat the construction for the other class-A Bianchi models.

The observables of the zeroth order theory are given by functions
$O^{0}=F(E^1,E^2,E^3,E^1 A_1-E^2 A_2,E^1 A_1-E^3 A_3)$ (if 
f one considers $F(E^1 A_1-E^2 A_2,E^1 A_1-E^3 A_3)$ these objects
are exact observables to all orders for the Bianchi I case).

As an example, we can start with an observable of the zeroth order
theory $O^{(0)}=E^1$.  The condition (\ref{seconcon}) translates into
the following partial differential equation
\begin{equation}
{\partial O^{(1)} \over \partial A_1} E^2 E^3
-E^1 E^2 A_2-
E^1 E^3 A_3+
\epsilon E^2 E^3+{\rm cyclic}
=0.
\end{equation}
This yields a first order correction, which modulo the
Hamiltonian constraint can be written as,
\begin{equation}
O^{(1)}_1 = -{A_1 \over E^2 E^3}
\left(-E^1 E^2 A_2+(E^1)^2 A_1-E^1 E^3 A_3 \right),
\end{equation}
up to an arbitrary additive function $F(E^1,E^2,E^3,E^1 A_1-E^2
A_2,E^1 A_1-E^3 A_3)$ that solves the homogeneous part of the
equation. Expressions for higher orders can easily be generated using
computer algebra, we do not list them here for reasons of space. The
behavior of the approximate observables can be studied explicitly
using the exact solution \cite{swe} of the equations of motion (in the
Bianchi I case). One observes that, evaluated on the trajectories of
the solution, the coefficients in the power series expansion lose
their dependence on the cosmological constant as an expansion
parameter. Carefully choosing the arbitrary functions $F$ one can
define observables that are finite at the Big Bang and remain
approximately constant for a significant portion of the lifetime of
the (recollapsing) universe. The rate of change decreases if
one adds further terms in the expansion, exhibiting convergence,
although all information about $\Lambda$ has disappeared.
Therefore quite unexpectedly, the method produces solutions that are
accurate {\em irrespective of the value of the cosmological
constant}.

Summarizing, casting ordinary general relativity as a perturbative
theory in $\Lambda^{-1}$ and  starting with background theory in which
$\Lambda\rightarrow\infty$ allows to explicitly compute non-degenerate
physical quantum states and observables. The technique is applicable
in the theory in general and also in minisuperspace models. In the
general theory, the calculations are well defined and tractable, but
quite involved. This opens up a program for the quantization of the
gravitational field that is well defined computationally and allows
contact with the physical observables of the theory, a key ingredient
in the road to finding a theory with a semiclassical regime that makes
contact with ordinary low energy  physics.

We wish to thank Abhay Ashtekar and Karel Kucha\v{r} for discussions.
This work was supported in part by the National Science Foundation
under grants PHY-9423950, INT-9811610, PHY-9407194, research funds of
the Pennsylvania State University and the Eberly Family research fund
at PSU.  We acknowledge support of PEDECIBA.

\end{multicols}
\begin{figure}
\centerline{\psfig{figure=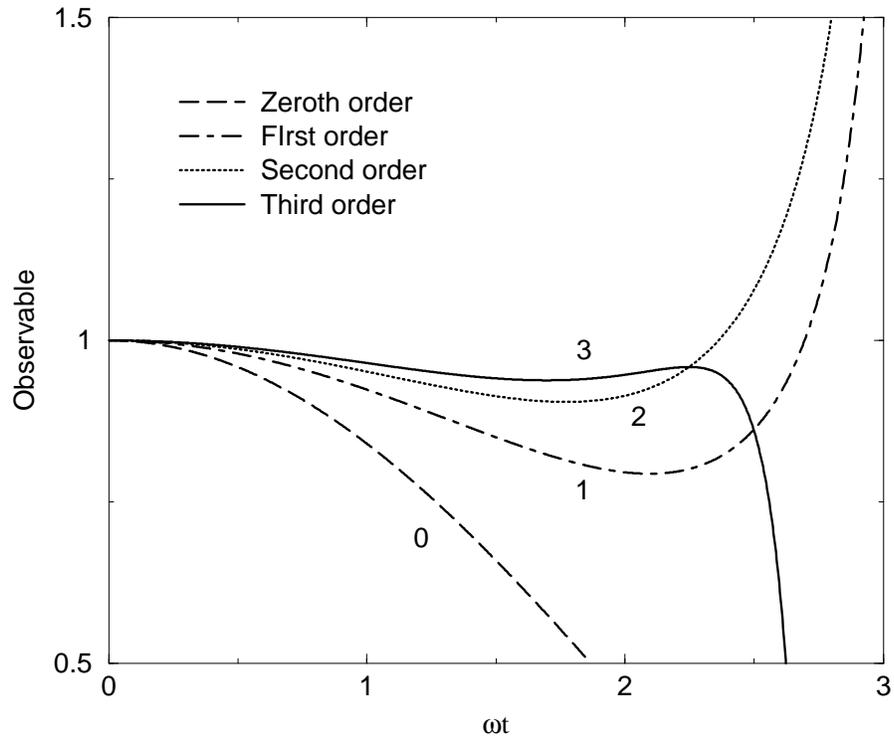,height=100mm}}
\caption{The perturbative observable evaluated for a Bianchi I
cosmology with negative cosmological constant. The time runs from
zero, the Big Bang, to $\pi$, the Big Crunch.  The four curves
correspond to better and better approximations.}
\end{figure}
\end{document}